\documentclass[12pt]{article}
\usepackage{amsmath,amssymb,amsfonts,amsthm}

\title{Entanglement criteria of two two-level atoms interacting with two coupled modes}

\author{H R Baghshahi$^{1,2,3}$, M J Faghihi$^{4,*}$ and M K Tavassoly$^{1,2}$ \\
 \footnotesize{$^1$ Atomic and Molecular Group, Faculty of Physics, Yazd University, Yazd, Iran} \\
 \footnotesize{$^2$ The Laboratory of Quantum Information Processing, Yazd University, Yazd, Iran} \\
 \footnotesize{$^3$ Department of Physics, Faculty of Science, Vali-e-Asr University of Rafsanjan, Rafsanjan, Iran} \\
 \footnotesize{$^4$ Physics and Photonics Department, Graduate University of Advanced Technology, Mahan, Kerman, Iran} \\
 \footnotesize{$^*$ E-mail: mj.faghihi@kgut.ac.ir}}

\begin{document}
\maketitle

 \newcommand{\norm}[1]{\left\Vert#1\right\Vert}
 \newcommand{\abs}[1]{\left\vert#1\right\vert}
 \newcommand{\set}[1]{\left\{#1\right\}}
 \newcommand{\R}{\mathbb R}
 \newcommand{\I}{\mathbb{I}}
 \newcommand{\C}{\mathbb C}
 \newcommand{\eps}{\varepsilon}
 \newcommand{\To}{\longrightarrow}
 \newcommand{\BX}{\mathbf{B}(X)}
 \newcommand{\HH}{\mathfrak{H}}
 \newcommand{\A}{\mathcal{A}}
 \newcommand{\D}{\mathcal{D}}
 \newcommand{\N}{\mathcal{N}}
 \newcommand{\x}{\mathcal{x}}
 \newcommand{\p}{\mathcal{p}}
 \newcommand{\la}{\lambda}
 \newcommand{\af}{a^{ }_F}
 \newcommand{\afd}{a^\dag_F}
 \newcommand{\afy}{a^{ }_{F^{-1}}}
 \newcommand{\afdy}{a^\dag_{F^{-1}}}
 \newcommand{\fn}{\phi^{ }_n}
 \newcommand{\HD}{\hat{\mathcal{H}}}
 \newcommand{\HDD}{\mathcal{H}}

 \begin{abstract}
In this paper, we study the interaction between two two-level atoms and two coupled modes of a quantized radiation field in the form of parametric frequency converter injecting within an optical cavity enclosed by a medium with Kerr nonlinearity. It is demonstrated that, by applying the Bogoliubov-Valatin canonical transformation, the introduced model is reduced to a well-known form of the generalized Jaynes-Cummings model. Then, under particular initial conditions which may be prepared for the atoms (in a coherent superposition of its ground and upper states) and the fields (in a standard coherent state), the time evolution of state vector of the entire system is analytically evaluated. In order to understand the degree of entanglement between subsystems (atom-field and atom-atom), the dynamics of entanglement through different measures, namely, von Neumann reduced entropy, concurrence and negativity is evaluated. In each case, the effects of Kerr nonlinearity and detuning parameter on the above criteria are numerically analyzed, in detail. It is illustrated that the amount of the degree of entanglement can be tuned by choosing the evolved parameters, appropriately.
 \end{abstract}



 \section{Introduction}\label{sec-intro}
Entanglement, manifesting an unbreakable quantum correlation between parts of a multipartite quantum system, is one of the most challenging features of quantum mechanics that is tightly related to its foundations, particularly quantum nonseparability and the violation of Bell's inequalities \cite{Einstein1935,Schrodinger1935,Bell1964}.
Also, quantum entanglement is considered as one of the key resources for quantum information science such as quantum computation and communication \cite{Bennett.DiVincenzo2000,Bengtsson2006}, quantum dense coding \cite{Li2002}, quantum teleportation \cite{Bennett.etal1993}, entanglement swapping \cite{Abdi.etal2012}, sensitive measurements \cite{Richter.Vogel2007}, and quantum telecloning \cite{Murao1999}.
However, apart from the entanglement, there exist other nonclassical correlations that have attracted great attention in this field of research. For instance, quantum discord \cite{Ollivier2001}, which expresses the basic aspect of classical bipartite states, is a criterion to characterize all nonclassical correlations. Anyway, the quantum entanglement is a fundamental concept in quantum information processing and plays a crucial role within new information technologies \cite{Benenti.etal2007}.
Nevertheless, some fundamental questions about entanglement, for instance, entanglement sudden death (ESD) and ESD revival (ESDR) remain open. In this direction, it has been shown that entanglement (of a two-qubit system) may decrease abruptly (also non-smoothly) and asymptotically tends to zero in a finite time. This phenomenon (ESD) is considered as a consequence of the presence of quantum/classical noise \cite{Yu2009}. It is worth noticing that, theoretical as well as experimental verification of ESD has been recently reported in literature \cite{Yu2004,Yu2006a,Laurat2007}. It is also valuable to mention that contrary to the occurrence of ESD, entanglement sudden birth (ESB) can suddenly be appeared \cite{Ficek2008}. \\
Altogether, manipulating and generating the entangled states are great attention.  For instance, it is recently reported that the entangled state can be used in quantum metrology \cite{Berrada2013} and in the violation of Bell's inequalities \cite{Gerry2009,Torlai2013}. In particular, the appearance of entanglement in the interaction between light and matter in cavity quantum electrodynamics (QED), as a simple way to produce the entangled states and as a candidate for the physical realization of quantum information processing, is of special interest. In this respect, the atom-field entangled states have been experimentally generated through interacting a single atom with a mesoscopic field in a high-Q microwave cavity \cite{Auffeves2003}. Also, based on cavity QED, quantum entanglement has been generated via sending two atoms being simultaneously in the cavity \cite{Zheng2000} or two atoms interacting successively with the cavity \cite{Phoenix1993}.\\
The Jaynes-Cummings model (JCM), as is well-known, is a fully quantum mechanical and exact soluble model which gives a pattern to describe the most basic and important interaction between light (a single-mode quantized electromagnetic field) and matter (a two-level atom) in the rotating wave approximation (RWA). Based on this model and also its generalizations which have been reported in the literature \cite{Prants2006,Tesfa2006,Sainz2007}, the atom-field interaction naturally leads to the quantum entangled state. In order to modify the JCM, various generalizations via using multi-level atoms \cite{Yadollahi.Tavassoly2011,Tavassoly.Yadollahi2012,Sahrai2013,Baghshahi.Tavassoly2014}, multi-mode fields \cite{Obada2013c}, multi-photon transitions \cite{Joshi2000,Bashkirov2008}, intensity-dependent coupling \cite{Buvzek1989,Fink2008}, Kerr nonlinearity \cite{Agarwal1989} and so on have been proposed throughout the recent five decades. For instance, a model describing the interaction between a general three-level atom and a bimodal cavity field has been carried out in \cite{Abdel-Aty2006}. An analytic solution for two two-level identical atoms interacting with a single-mode quantized radiation field containing
Stark shift has been studied in \cite{Hessian2011}. The case of the interaction between a $\Lambda$-type three-level atom and a single-mode field in an optical cavity surrounded by Kerr nonlinearity with intensity-dependent coupling has been studied \cite{Faghihi.Tavassoly2012}. The ability of the nonlinear JCM in generating a class of $SU(1,1)$ coherent states of the Gilmore-Perelomov type and also $SU(2)$ group was shown in \cite{Miry.Tavassoly2012}. In addition, as a result of a system in which a two-level atom interacts alternatively with a dispersive quantized cavity field and a resonant classical field, a theoretical scheme from which the nonlinear elliptical states can be generated has been newly proposed \cite{Miry.etal2013}. Recently, the nonlinear interaction between a $\Lambda$-type three-level atom and a two-mode cavity field in the presence of a Kerr medium and its deformed counterpart \cite{Honarasa.Tavassoly2012} has been investigated \cite{Faghihi.etal2013,Faghihi.Tavassoly2013a}. In addition, a model for a moving three-level JCM in the presence of intensity-dependent coupling has been  proposed \cite{Faghihi.Tavassoly2013}. More recently, some of the well-known nonclassicality features of the nonlinear (intensity-dependent) interaction between two two-level atoms and a single-mode binomial field has been reported \cite{Hekmatara2014}. Moreover, the full nonlinear $k$-photon JCM in the presence of intensity-dependent Stark shift and deformed Kerr medium \cite{Honarasa.Tavassoly2012} including time-varying atom-field coupling has been discussed \cite{Baghshahi2014}.
In the above-mentioned studies, showing only a few of the papers concerned with the JCM and its extensions, in particular the degree of entanglement (DEM) is evaluated via von Neumann reduced entropy besides some of the other nonclassicality features. \\
From another perspective of this field of research and also in direct relation to the present work, one may consider two coupled quantized fields jointly entering a high-Q bichromatic cavity \cite{Abdalla2005,Abdel-Aty.etal2009,Faghihi2014,Faghihi2014a}. In this way, the atom can interact with each field mode individually as well as both fields. Moreover, in a particular case, the atom can interact with two coupled fields in the form of frequency convertor type which are injected within the cavity \cite{Khalil2006}. \\
In this paper, we intend to study the interaction between two two-level atoms (which are considered to be initially in a coherent superposition of its ground and excite state) with two coupled quantized fields in the form of parametric converter type injected within an optical cavity with a centrosymmetric nonlinear (Kerr) medium. After considering all existing interactions appropriately in the Hamiltonian model of the entire atom-field system, in order to be able to solve the model, we have to apply a particular canonical transformation, namely Bogoliubov-Valatin canonical transformation, to reduce our complicated model to a form that can be analytically analyzed \cite{Faghihi.etal2013,Faghihi.Tavassoly2013a}. As a result, the explicit form of the state vector of the whole system is exactly obtained by using the time-dependent Schr\"{o}dinger equation. So, briefly, the main goal of this paper is to investigate individually and simultaneously the effects of Kerr medium (containing self- and cross-action) and detuning parameter on the entanglement dynamics. To achieve this purpose, the amount of the DEM between subsystems (atom-field and atom-atom) by using the von Neumann reduced entropy, concurrence and negativity is studied, in detail.
To clarify our motivations of this paper, it is valuable to give a few words on the notability of the entanglement between the fields. Even though, the importance of quantum entanglement has been previously declared, it is instructive to point out the entangled state due to the fields which may be utilized as input data for new researches/observations such as in the field of quantum computation. It is shown that two electromagnetic field modes of a cavity can be applied as a universal quantum logic gate \cite{Sanders1992,Sanders1992a}.\\
The remainder of paper is organized as follows: In the next section, by applying the canonical transformation, the state vector of the whole system is exactly obtained. In section 3, in order to understand the DEM between subsystems such as atom-field and atom-atom, we deal with von Neumann reduced entropy, concurrence and negativity. Finally, section 4 contains a summary and concluding remarks.
%
 \section{The model and its solution}\label{model}
%
The main goal of this section is to obtain the state vector of the system, in which two two-level atoms interact simultaneously with two coupled fields in the form of parametric converter type in an optical cavity involving Kerr nonlinearity with detuning parameter.
Based on the fundamentals of quantum mechanics, all necessary information about any physical (quantum) system is hidden in its wave function. This may be obtained after an exact view on the existing interactions between subsystems is truly achieved.
So, let us assume a model in which the two quantized radiation fields oscillating with frequencies $\Omega_{1}$ and $\Omega_{2}$ interact simultaneously with two two-level atomic system in the optical cavity which is surrounded by the Kerr medium. Also, considering the centrosymmetric nonlinear medium, self- and cross-action of the Kerr nonlinearity should properly be taken into account (see figure \ref{Diagram}). Anyway, the Hamiltonian containing all existing interactions which describes the dynamics of the introduced physical system in the RWA can be written as ($\hbar=c=1$):
\begin{equation} \label{1}
\hat{H}=\hat{H}_{A}+\hat{H}_{F}+\hat{H}_{AF},
\end{equation}
where the atomic and field parts of the Hamiltonian are given by
 \begin{eqnarray} \label{2}
\hat{H}_{A} &=& \frac{1}{2}\sum_{ j = 1 }^{2}\omega_{j}\hat{\sigma}^{(j)}_{z}, \nonumber\\
\hat{H}_{F} &=& \sum_{ j = 1 }^{2} \left( \Omega_{j} \hat{a}_{j}^{\dag} \hat{a}_{j} + \chi_{j} \hat{a}_{j}^{\dag 2} \hat{a}_{j}^{2} \right)
+ \chi_{12} \hat{a}_{1}^{\dag}\hat{a}_{1}\hat{a}_{2}^{\dag} \hat{a}_{2}
+ \lambda_{12} \left( \hat{a}_{1}^{\dag}\hat{a}_{2}+\hat{a}_{1}\hat{a}_{2}^{\dag} \right),
\end{eqnarray}
and the atom-field interaction reads as
\begin{equation} \label{3}
\hat{H}_{AF} = i \left( \hat{a}_{1}^{\dag}\hat{a}_{2}-\hat{a}_{1}\hat{a}_{2}^{\dag} \right) \sum_{ j = 1 }^{2}\lambda_{j} \left( \hat{\sigma}_{+}^{(j)}+\hat{\sigma}_{-}^{(j)} \right),
\end{equation}
where $\omega_{j}$, $\lambda_{12}$ and $\lambda_{j} \; (j=1,2)$ are respectively the frequency of atomic transition, the $j$th mode frequency, field-field coupling and atom-field coupling. In addition, $\chi_{j}, (j=1,2)$ and $\chi_{12}$ are referred to as the cubic susceptibility of the medium; $\chi_{j}$ shows the Kerr self-action for mode $j$ and $\chi_{12}$ is related to the Kerr cross-action process. The two two-level atoms are described by the atomic pseudospin operators $\hat{\sigma}_{z}^{(j)}$ and $\hat{\sigma}_{\pm}^{(j)} (j=1,2)$, and the bosonic operators $\hat{a}_{j}^{\dag}$ and $\hat{a}_{j}, (j=1,2)$ are the field creation and annihilation operators, respectively. \\
Now, in order to analyze the dynamics of the considered quantum system described by the Hamiltonian in equation (\ref{1})), there exist three different but equivalent methods, namely, probability amplitudes, Heisenberg operators and the unitary time evolution operator approaches \cite{Scully.Zubairy1997}.
For the introduced model we prefer to use the probability amplitude method. But before this, it is convenient to simplify the complicated system by applying the following canonical transformations
 \begin{equation} \label{4}
\hat{a}_{1}=\hat{b}_{1}\cos\theta+\hat{b}_{2}\sin\theta,    \hspace{0.75cm}
\hat{a}_{2}=\hat{b}_{2}\cos\theta-\hat{b}_{1}\sin\theta,
\end{equation}
 which are the well-known Bogoliubov-Valatin transformations \cite{Svozil1990,Jauregui1992} and have been introduced in the context of the  Bardeen-Cooper-Schrieffer model of superconductivity \cite{Schrieffer1964}. In these transformations, the operators $\hat{b}_{j}$ ($\hat{b}_{j}^{\dag}$), $j=1,2$, are the bosonic annihilation (creation) operators having similar properties to $\hat{a}_{j}$ ($\hat{a}_{j}^{\dag}$), respectively. Also, parameter $\theta$ is the rotation angle which will be determined later. It is worth to mention that, under these transformations the total number of photons is invariant, that is, $\hat{a}_{1}^{\dag}\hat{a}_{1}+\hat{a}_{2}^{\dag}\hat{a}_{2}=\hat{b}_{1}^{\dag}\hat{b}_{1}+\hat{b}_{2}^{\dag}\hat{b}_{2}$. By inserting equation (\ref{4}) into Hamiltonian (\ref{1}) and after some straightforward calculations, one may obtain the transformed Hamiltonian in the form
 \begin{eqnarray}  \label{5}
\hat{H}&=&\sum_{j=1}^{2} \bigg[\frac{1}{2} \omega_{j} \hat{\sigma}^{(j)}_{z} +
\bar{\Omega}_{j}\hat{b}_{j}^{\dag}\hat{b}_{j}+\bar{\chi }\hat{b}_{j}^{\dag 2}\hat{b}_{j}^{2} \nonumber \\
&+& i \lambda_{j} \left(\hat{b}_{1}^{\dag}\hat{b}_{2}{\sigma}_{+}^{(j)} - \hat{b}_{1}\hat{b}_{2}^{\dag}\hat{\sigma}_{-}^{(j)}\right) \bigg]
+\sum_{j \neq k = 1}^{2}\bar{\chi}\hat{b}_{j}^{\dag}\hat{b}_{j}\hat{b}_{k}^{\dag}\hat{b}_{k}.
\end{eqnarray}
Looking deeply at the latter relations implies the fact that, the canonical transformations in equation (\ref{4}) retain the invariance of the Kerr nonlinearities so long as the relation $\chi_{1} = \chi_{2} = \chi = \chi_{12}/ 2$ is satisfied. Meanwhile, we have also defined
\begin{eqnarray} \label{6}
&\bar{\Omega}_{1}&=\Omega_{1}\cos^{2}\theta+\Omega_{2}\sin^{2}\theta-\lambda_{12}\sin2\theta,\nonumber \\
&\bar{\Omega}_{2}&=\Omega_{1}\sin^{2}\theta+\Omega_{2}\cos^{2}\theta+\lambda_{12}\sin2\theta,
\end{eqnarray}
in which the rotation angle $\theta$ is still unknown and should be determined. For this purpose, the evanescent wave terms from the Hamiltonian related to the fields and field-field interaction should be avoided. Therefore, one may set the particular choice of angle $\theta$ as
\begin{equation} \label{7}
\theta = \frac{1}{2}\tan^{-1}\left(\frac{2\lambda_{12}}{\Omega_{1} - \Omega_{2}}\right).
\end{equation}
It should be noted that in obtaining equation (\ref{5}), we have applied the RWA with respect to the rotated operators $\hat{b}_{i}$ and $\hat{b}_{i}^{\dag}$. It is suitable to work in the interaction picture with the Hamiltonian $V_{I}(t)$ as follows
\begin{eqnarray}  \label{8}
\hat{V_{I}}(t)&=& i \sum_{j=1}^{2}\lambda_{j} \left(\hat{b}_{1}^{\dag}\hat{b}_{2}{\sigma}_{+}^{(j)} e^{i\Delta_{j}t}
 - \hat{b}_{1}\hat{b}_{2}^{\dag}\hat{\sigma}_{-}^{(j)}e^{-i\Delta_{j}t} \right)  \nonumber \\
&+& \sum_{ j = 1}^{2}\bar{\chi}\hat{b}_{j}^{\dag 2} \hat{b}_{j}^{2}
+ \sum_{j \neq k=1}^{2}\bar{\chi}\hat{b}_{j}^{\dag}\hat{b}_{j}\hat{b}_{k}^{\dag}\hat{b}_{k},
\end{eqnarray}
where $\Delta_{j}=\omega_{j}-(\bar{\Omega}_{2}-\bar{\Omega}_{1})$ is the detuning parameter.
To obtain the explicit form of the wave function of the whole system, we solve the time-dependent Schr\"{o}dinger equation
$i\frac{\partial}{\partial t}|\psi(t)\rangle=\hat{V}_{I}(t)|\psi(t)\rangle$.
For the assumed system, the wave function at any time $t$ can be written in the following form:
\begin{eqnarray} \label{9}
|\psi(t)\rangle&=&\sum_{n_{1}=0}^{\infty}\sum_{n_{2}=0}^{\infty} q_{n_{1}}q_{n_{2}} \bigg [ A(n_{1}+2,n_{2},t)|e_{1},e_{2},n_{1}+2,n_{2}\rangle \nonumber\\
&+&B(n_{1}+1,n_{2}+1,t)|e_{1},g_{2},n_{1}+1,n_{2}+1\rangle \nonumber \\
&+&C(n_{1}+1,n_{2}+1,t)|g_{1},e_{2},n_{1}+1,n_{2}+1\rangle \nonumber \\
&+&D(n_{1},n_{2}+2,t)|g_{1},g_{2},n_{1},n_{2} + 2\rangle \bigg],
 \end{eqnarray}
where $q_{n_{1}}$ and $q_{n_{2}}$ are the probability amplitudes of the initial field state of the radiation field of the cavity field. Also, $A$, $B$, $C$ and $D$ are the atomic probability amplitudes which have to be determined.
For this purpose, by taking the probability amplitude technique into account, one may arrive at the following coupled differential equations for the probability amplitudes:
\begin{eqnarray} \label{10}
\dot{A}&=& f_{1}^{(2)} e^{i \Delta_{2} t} B+f_{1}^{(1)} e^{i \Delta_{1} t} C-iV_{1} A,    \nonumber\\
\dot{B}&=& -f_{1}^{(2)} e^{-i\Delta_{2} t}A+ f_{2}^{(1)}e^{i \Delta_{1} t} D-i V_{2} B,   \nonumber\\
\dot{C}&=& -f_{1}^{(1)} e^{-i\Delta_{1} t}A+f_{2}^{(2)} e^{i \Delta_{2} t} D-i V_{2} C,   \nonumber\\
\dot{D}&=& - f_{2}^{(1)} e^{-i \Delta_{1} t}B- f_{2}^{(2)} e^{-i \Delta_{2} t}C-i V_{3} D,
 \end{eqnarray}
where
\begin{eqnarray} \label{11}
f_{1}^{(j)}&=&\lambda_{j}\sqrt{(n_{1}+2)(n_{2}+1)},\nonumber\\
f_{2}^{(j)}&=&\lambda_{j}\sqrt{(n_{1}+1)(n_{2}+2)},\hspace{0.5cm} j=1,2, \nonumber\\
V_{1}&=&\chi \big[(n_{1}+2)(n_{1}+1)+n_{2}(n_{2}-1)+2(n_{1}+2)n_{2} \big],   \nonumber \\
V_{2}&=&\chi \big[n_{1}(n_{1}+1)+n_{2}(n_{2}+1)+2(n_{1}+1)(n_{2}+1)\big],    \nonumber \\
V_{3}&=&\chi \big[n_{1}(n_{1}-1)+(n_{2}+1)(n_{2}+2)+2n_{1}(n_{2}+2)\big].     \nonumber
 \end{eqnarray}
Now, in order to obtain a formalism close to the experimental situations, we consider the atoms to be identical, i.e., $f_{1}^{(1)}=f_{1}^{(2)}=f_{1}$, $f_{2}^{(1)}=f_{2}^{(2)}=f_{2}$ and $\Delta_{1}=\Delta_{2}=\Delta$. As a result, the coefficients $B$ and $C$ are the same and so the four coupled differential equations in (\ref{10}) are reduced to three ones which are given by
 \begin{eqnarray} \label{12}
\dot{A}&=& 2 f_{1} e^{i \Delta t} B-iV_{1} A, \nonumber\\
\dot{B}&=& -f_{1} e^{-i\Delta t}A+f_{2} e^{i \Delta t} D-i V_{2} B, \nonumber\\
\dot{D}&=& -2 f_{2} e^{-i \Delta t}B-i V_{3} D.
 \end{eqnarray}
By inserting $D = e^{i \mu t}$ in equation (\ref{12})  we obtain the third-order equation
\begin{equation}\label{13}
\mu^{3}+x_{1}\mu^{2}+x_{2}\mu+x_{3}=0,
\end{equation}
where
\begin{eqnarray} \label{14}
x_{1} &=& 3\Delta+V_{1}+V_{2}+V_{3}, \nonumber\\
x_{2} &=& -2(f_{1}^{2}+f_{2}^{2})+(2\Delta+V_{1})(\Delta+V_{2}) + (3\Delta+V_{1}+V_{2})V_{3},   \nonumber \\
x_{3} &=& -2f_{2}^{2}(2\Delta+V_{1})+(-2f_{1}^{2}+(2\Delta  + V_{1})(\Delta+V_{2}))V_{3}.
  \end{eqnarray}
 As presented in \cite{Faghihi.Tavassoly2012}, equation (\ref{13}) has generally three different roots that can be expressed in the form
 \begin{eqnarray} \label{15}
 \mu_{m}&=&-\frac{1}{3}x_{1}+\frac{2}{3}\sqrt{x_{1}^{2}-3x_{2}}\cos \left(\phi+\frac{2}{3}(m-1)\pi\right), \nonumber \\
 m &=& 1,2,3,  \phi = \frac{1}{3}\cos^{-1}\left[ \frac{9 x_{1}x_{2}-2x_{1}^{3}-27x_{3}}{2(x_{1}^{2}-3x_{2})^{3/2}}\right].
  \end{eqnarray}
  Consequently, $D$ can be considered as a linear combination of $e^{i \mu_{m} t}$ as follows
  \begin{equation} \label{16}
  D = \sum_{m=0}^{3}b_{m}e^{i \mu_{m} t}.
   \end{equation}
At last, by inserting (\ref{16}) into equations (\ref{12}) and after some lengthy but simple manipulations, the probability amplitudes $A$, $B$, $C$ and $D$ (specifying the explicit form of the state vector of whole system) may be found in the form
 \begin{eqnarray} \label{17}
 A(n_{1}+2,n_{2},t)&=&\frac{e^{2i\Delta t}}{2f_{1}f_{2}}\sum_{m=0}^{3} \bigg(2f_{2}^{2}-(\mu_{m}+V_{3})
 (\mu_{m}+V_{2}+\Delta) \bigg) b_{m}e^{i\mu_{m}t}, \nonumber \\
 B(n_{1}+1,n_{2}+1,t)&=&C(n_{1}+1,n_{2}+1,t) = \frac{-ie^{i\Delta t}}{2f_{2}}\sum_{m=0}^{3}(\mu_{m}+V_{3})  b_{m}e^{i\mu_{m}t}, \nonumber\\
 D(n_{1},n_{2}+2,t)&=&\sum_{m=0}^{3}b_{m}e^{i\mu_{m}t},
  \end{eqnarray}
where $b_{m}$ should be determined via the initial condition of the atoms. So, let us consider the atoms enter the cavity in the general coherent superposition state of the ground and the exited states, that is,
\begin{equation} \label{19}
|\psi_{\mathrm{atoms}}(t=0)\rangle=\cos(\beta/2)|e_{1},e_{1}\rangle+\sin(\beta/2)|g_{1},g_{2}\rangle,
\end{equation}
where $0\leq\beta\leq\pi$, i.e. $A(0)=\cos(\beta/2)$, $B(0) = 0 =C(0)$ and $D(0)=\sin(\beta/2)$. Then, the following relation may be obtained:
\begin{equation} \label{20}
b_{m}=\frac{-2\cos(\beta/2)f_{1}f_{2}+\sin(\beta/2) (2f_{2}^{2}+(V_{3}+\mu_{k})(V_{3}+\mu_{l}))}{\mu_{mk}\mu_{ml}},\hspace{2cm}m\neq k \neq l,
\end{equation}
where $\mu_{mk} = \mu_{m} - \mu_{k}$. Consequently, the probability amplitudes $A$, $B$, $C$ and $D$ are exactly derived. \\
It is now worthwhile to mention that, in order to study the dynamics of entanglement, arbitrary amplitudes of the initial states of the field such as number, phase, coherent or squeezed state can be considered. However, since the coherent state (the laser field far above the threshold condition \cite{Scully.Zubairy1997}) is more accessible than other typical field states, we shall consider the fields to be initially in the coherent state
\begin{equation} \label{18}
|\alpha_{i}\rangle=\sum_{n_{1}=0}^{\infty}\sum_{n_{2}=0}^{\infty}q_{n_{1}}q_{n_{2}}|n_{1},n_{2}\rangle,\hspace{0.1cm}q_{n_{i}}=\exp(\frac{-|\alpha_{i}|^{2}}{2})\frac{\alpha_{i}^{n_{i}}}{\sqrt{n_{i}!}},
\end{equation}
where $|\alpha_{1}|^{2}$ and $|\alpha_{2}|^{2}$ show the initial mean photon number of mode $1$ and $2$, respectively.
Accordingly, the exact form of the wave function $ | \psi (t) \rangle $ introduced in (\ref{9}) is explicitly obtained. In the next section, we shall investigate the DEM between different subsystems by using the proper measures.
%
\section{Entanglement criteria}
%
Quantifying the entanglement is achieved through suitable measures that are well justified and mathematically tractable \cite{Horodecki2009}. Some of the measures that qualify the necessary conditions are, for instance, entanglement of formation and distillation \cite{Wootters1998}, negativity \cite{Vidal2002}, von Neumann entropy and relative entropy \cite{Vedral1997} and concurrence \cite{Uhlmann2000}. This section is allocated to study the von Neumann reduced entropy, concurrence and negativity, in order to obtain the DEM between subsystems. In each case, the effect of Kerr nonlinearity, as well as detuning parameter, is numerically examined.
%
\subsection{ von Neumann entropy}
%
The quantum entropy (quantum mutual information) is a useful criterion to evaluate the DEM. In other words, the time evolution of the entropy of the field or the atom reflects the time evolution of the DEM between subsystems. Meanwhile, the reduced von Neumann entropy, as a measure of the DEM, satisfy the general conditions consist of Schmidt decomposition, local invariance, continuity and additivity \cite{Nielsen.Chuang2010}. Before obtaining the reduced entropy of the field and the atom, it is valuable to pay attention to the important theorem of Araki and Leib \cite{Araki.Lieb1970}. According to this theorem, in a bipartite quantum system, the system and subsystem entropies at any time $t$ are limited by the triangle inequality $|S_{A}(t)-S_{F}(t)|\leq S\leq|S_{A}(t)+S_{F}(t)|$,
where the subscripts `A' and `F' refer to the atom and the field, respectively, and the total entropy of the atom-field system is denoted by $S$. As a result, if at the initial time the field and the atom are in pure states, the total entropy of the system is zero and remains constant. This means that if the system is initially prepared in a pure state (as we have considered), at any time $t >0$, the reduced entropies of the two subsystems (atom and field) are identical, that is, $S_{A}(t)=S_{F}(t)$ \cite{Phoenix.Knight1990,Barnett.Phoenix1991}. Therefore, both atomic and field entropies are equivalent measures of the entanglement. So, we concentrate on the evolution of the atomic entropy against time to obtain the DEM. The reduced entropy of the atom (field) according to the von Neumann entropy is defined through the corresponding reduced density operator by
\begin{equation} \label{22}
S_{A(F)}(t)=-\mathrm{Tr}_{A(F)}(\hat{\rho}_{A(F)}(t)\ln\hat{\rho}_{A(F)}(t)),
\end{equation}
where $\hat{\rho}_{A(F)}(t)=\mathrm{Tr}_{F(A)}(|\psi(t)\rangle \langle\psi(t)|)$ is the reduced density operator of the atom (field).
Following the procedure of \cite{Faghihi.etal2013}, one may express the entropy of the field/atom by the following relation
\begin{equation} \label{25}
\mathrm{DEM}(t) = S_{F}(t) = S_{A}(t)=- \sum_{i=1}^{4}\xi_{i} \ln\xi_{i},
\end{equation}
where $\xi_{i}$, the eigenvalues of the reduced atomic density matrix are given by Cardano's method as \cite{Childs2009}
\begin{eqnarray}
\xi_{j} &=& -\frac{1}{3}\zeta_{1}+\frac{2}{3}\sqrt{\zeta_{1}^{2}-3\zeta_{2}} \cos \left(\varpi+\frac{2}{3}(j-1)\pi \right), \;\;\;\;\;\;\;  j = 1,2,3, \nonumber \\
\xi_{4} &=& 0,
 \end{eqnarray}
with
\begin{equation} \label{27}
\varpi=\frac{1}{3}\cos^{-1}\left[ \frac{9 \zeta_{1}\zeta_{2}-2 \zeta_{1}^{3}-27\zeta_{3}}{2(\zeta_{1}^{2}-3\zeta_{2})^{3/2}}\right],
\end{equation}
and
\begin{eqnarray}\label{28}
\zeta_{1} &=& -\rho_{11}-2\rho_{22}-\rho_{44}, \nonumber\\
\zeta_{2} &=& -2\rho_{12}\rho_{21}-\rho_{14}\rho_{41}-2\rho_{24}\rho_{42}+2\rho_{22}\rho_{44} + \rho_{11}(2\rho_{22}+\rho_{44}), \nonumber \\
\zeta_{3} &=& 2\rho_{14}(\rho_{22}\rho_{41}-\rho_{21}\rho_{42})+\rho_{12}(\rho_{21}\rho_{44}-\rho_{24}\rho_{41}) + \rho_{11}(\rho_{24}\rho_{42}
-\rho_{22}\rho_{44}),
\end{eqnarray}
with
\begin{eqnarray}\label{281}
\rho_{ij}(t)&=&\sum_{n=0}^{\infty}\sum_{m=0}^{\infty} \langle n,m,i|\psi(t)\rangle \langle \psi(t)| n,m, j\rangle, \hspace{1cm}
i,j = 1,2,3,4.
\end{eqnarray}
Equation (\ref{25}) indicates the time evolution of the field/atom entropy. In addition, it is fruitful to notice that, by this equation, the DEM between the atoms and fields is also determined, i.e. the subsystems are disentangled (the system of atom-field is separable) if equation (\ref{25}) tends to zero. \\
Figure \ref{Entanglement} shows the time evolution of the field entropy against the scaled time $\tau$ for the initial mean number of photons fixed at $|\alpha_{1}|^{2} = 10 = |\alpha_{2}|^{2}$.
Figure \ref{Entanglement}(a) shows that in the resonance case and in the absence of Kerr nonlinearity $(\chi=0, \Delta=0)$, the DEM has a random behaviour. By entering the effect of detuning parameter in figure \ref{Entanglement}(b), the temporal behaviour of the DEM with fast oscillatory is observed and the maximum amount of DEM is increased. From figure \ref{Entanglement}(c) where the effect of Kerr nonlinearity is studied, it is seen that, the amount of DEM in the presence of Kerr nonlinearity is drastically reduced. In figure \ref{Entanglement}(d), the effect of all considered parameters is examined. According to this figure and comparing with figure \ref{Entanglement}(c), it is observed that, in the presence of Kerr medium, detuning parameter may enhance the DEM.

%
\subsection{Concurrence}
%
The concurrence presented by Hill and Wootters \cite{Hill1997,Wootters1998} is a suitable measure of the entanglement of any state of two qubits, mixed or pure. This quantity is described by using the Pauli spin matrix $\sigma_{y}$ as a spin flip operator. Rungta \textit{et al} \cite{Rungta2001} generalized the spin-flip superoperator to a `universal inverter' which acts on quantum systems with arbitrary dimension. The authors showed that, for a pure state $ | \psi \rangle $ on $(M \times N)$-dimensional Hilbert space $ \mathfrak{R} = \mathfrak{R}_{M} \otimes \mathfrak{R}_{N} $, the concurrence (that they called `\textit{I concurrence}') can be defined as follows \cite{Abdel-Aty2006}
  \begin{equation}\label{3311}
C(|\psi\rangle)=\sqrt{2(\langle |\psi| \psi \rangle|^{2} - \mathrm{Tr}(\rho_{N}^{2}))},
\end{equation}
 in which $\hat{\rho}_{N} = \mathrm{Tr}_{M}(|\psi\rangle \langle\psi(t)|)$ is the reduced density operator of the subsystem with dimension $N$ where $\mathrm{Tr}_{M}$ is the partial trace over $\mathfrak{R}_{M}$.
 It is noteworthy to mention that, the concurrence varies between $0$ for separable state and $\sqrt{2(N-1)/N}$ for maximally entangled state.
 Here, we calculate the concurrence to understand the DEM between the atoms and the fields. To achieve this purpose, it is convenient to rewrite the atomic reduced density operator in the form
 \begin{equation}\label{3312}
 \hat{\rho}_{A} = \sum_{i,j=1,2,3,4} \rho_{ii}|i \rangle \langle i| + \sum_{i,j=1,2,3,4, i \neq j} \rho_{ij} |i \rangle \langle j |,
\end{equation}
where $ \rho_{ij}=\langle i| \hat{\rho}_{A}(t) | j \rangle$. Hence, the concurrence defined in (\ref{3311}) results in
  \begin{equation}\label{3313}
C=\sqrt{2 \sum_{i,j=1,2,3,4,i\neq j}(\rho_{ii} \rho_{jj}-\rho_{ij} \rho_{ji})}.
\end{equation}
In figure \ref{Concurrence}, we have plotted the concurrence versus $\tau$ with the same parameters similar to figure \ref{Entanglement}. From figure \ref{Concurrence}(a) whereas detuning parameter and Kerr nonlinearity are neglected, a random behavior is observed. Figure \ref{Concurrence}(b) has been plotted in off resonance case and in the absence of Kerr effect. It is obvious from this figure that, when the time goes up, the DEM is increased. To investigate the effect of Kerr medium in the resonance and nonresonance case, we have depicted figures \ref{Concurrence}(c) and \ref{Concurrence}(d), respectively.
From figure \ref{Concurrence}(c) where the effect of Kerr nonlinearity is studied, it is observed  that, Kerr medium ruins the entanglement between the atoms and the fields. Also, from figure \ref{Concurrence}(d), it is seen that, in the presence of Kerr medium, the DEM may be improved when the detuning parameter is present.
Interestingly, comparing the presented results of figures \ref{Entanglement} and \ref{Concurrence} implies the fact that, dynamics of entanglement via von Neumann entropy and concurrence are qualitatively the same, as is expected.
This may be considered as a check point for the accuracy of our numerical calculation regarding the DEM between the atoms and the fields. It is also remarkable to declare that, the sensitivity of concurrence is clearly more than von Neumann entropy.
%
\subsection{Negativity}
%
In this section, we are mainly interested in analyzing the effect of the detuning parameter and Kerr nonlinearity on the entanglement between the atoms (atom-atom entanglement). It is valuable to mention that, two-atom entangled states have experimentally been reported by considering ultra cold trapped ions \cite{DeVoe1996} and cavity QED schemes \cite{Hagley1997}. The negativity is an appropriate and a good computable measure of this type of entanglement.
The concept of negativity is referred to the Peres-Horodecki condition for the separability of a state \cite{Peres1996,Horodecki1996}. According to this quantity, a state is entangled if one or more of the eigenvalues of partial transposition matrix is negative. The Peres criterion is necessary and sufficient for the  ($2 \times 2$)- and ($2 \times 3$)-dimensional states, and is only sufficient for systems with higher dimensions.
The negativity for a bipartite quantum system with $(2 \times 2)$- and $(2 \times 3)$-dimensional Hilbert space described by density matrix $\hat{\rho}$, is given by \cite{Vidal2002}
 \begin{equation}\label{331}
\mathcal{N}(\hat{\rho}) = \frac{||\hat{\rho}^{T_{B}}||_{1}-1}{2},
\end{equation}
  where $\hat{\rho}^{T_{B}}$  is the partial transpose of a state $\hat{\rho}$ (with respect to subsystem $B$), and $||\hat{\rho}^{T_{B}}||_{1}$ represents the trace norm of the operator $\hat{\rho}^{T_{B}}$.
 It ought to be mentioned that, the trace norm of any operator $\hat{O}$  is defined by $||\hat{O}||_{1}=\mathrm{Tr}\sqrt{\hat{O}^{\dag} \hat{O}}$ which is equal to the sum of the absolute values of the eigenvalues of $\hat{O}$, when the operator $\hat{O}$ is Hermitian.
 Also, It may be noted that, the quantity defined in (\ref{331}) is zero for separable states and strictly less than one for maximally entangled states.
 Also, it is valuable to notice that for a system with arbitrary dimensions, $d \times d'$ ($ d \leq d' $), the negativity can be expressed as \cite{Lee2003}
 \begin{equation}\label{3312}
\mathcal{N}(\hat{\rho}) = \frac{||\hat{\rho}^{T_{B}}||_{1} - 1}{d - 1},
\end{equation}
where $\hat{\rho}^{T_{B}}$  is the partial transpose of a state $\hat{\rho}$ in $ d \otimes d' $ quantum system. It is instructive to state that, this quantity gets $1$ for any pure maximally entangled state (such as one of the Bell states).
The matrix $\hat{\rho}$ has positive eigenvalues and so $\mathrm{Tr}(\hat{\rho})=1$. Also, for the partial transpose of this matrix we have $\mathrm{Tr}(\hat{\rho}^{T_{B}})=1$, too. Since the partial transpose of density operator might have the negative eigenvalues, the trace norm of $\hat{\rho}^{T_{B}}$ can be written in the following form \cite{Akhtarshenas2007}
\begin{equation}\label{332}
||\hat{\rho}^{T_{B}}||_{1}=\sum_{i}|\mu_{i}|=\sum_{i}\mu_{i}-2\sum_{i}\mu^{neg}_{i}=1-2\sum_{i}\mu^{neg}_{i},
\end{equation}
where $\mu_{i}$ and $\mu^{neg}_{i}$ correspond to the positive and negative eigenvalues of $\hat{\rho}^{T_{B}}$, respectively.
Here, we are going to evaluate the DEM between two two-level atoms with density operator $\hat{\rho}_{A}$, by using the negativity with emphasis on the fact that, for considered atomic subsystems $d = d' = 2$ ($(2 \times 2)$-dimensional Hilbert space). Considering the equation (\ref{3312}), the negativity reads as
\begin{equation}\label{333}
\mathcal{N} = - 2 \sum_{i}\mu^{neg}_{i},
\end{equation}
or equivalently $\mathcal{N} = \max(0,-2\sum_{i}\mu^{neg}_{i})$ \cite{Hessian2011}.
Consequently, we need only to calculate eigenvalues of $\hat{\rho}_{A}^{T_{A_{2}}}$ (partial transpose of reduced atomic density matrix with respect to the second atom) to arrive one at the DEM between the two atoms. \\
Our results presented in figure \ref{Negativity} exhibit the time evolution of the negativity against the scaled time $\tau$ for different chosen parameters as in figure \ref{Entanglement}. From figure \ref{Negativity}(a) which corresponds to the resonance condition and no Kerr medium, it is observed that the behavior of DEM between two atoms is random. Looking at figure \ref{Negativity}(b) indicates that in the presence of detuning parameter, the amount of DEM between the atoms is reduced. In figure \ref{Negativity}(c) in which the effect of Kerr medium is considered, the DEM strongly is descended.
In figure \ref{Negativity}(d), where all parameters (Kerr medium and detuning parameter) are considered, it is seen that, the negativity is lost when the time proceeds.
%
 \section{Summary and conclusion}\label{summary}
%
In this paper, we have studied two two-level atoms interacting with two coupled quantized fields in the form frequency converter type which are injected simultaneously within a bichromatic cavity enclosed by the centrosymmetric Kerr medium in the presence of detuning parameter. Next, by suitably considering all existing interactions, we have applied the Bogoliubov-Valatin canonical transformation, in order to reduce our complicated model to the usual form of the generalized JCM. After obtaining analytically the exact form of the state vector of the whole system, the effect of Kerr nonlinearity and detuning parameter on some of the well-known entanglement criteria has been individually and simultaneously examined. To reach this goal, the time evolution of different types of entanglement measures, consisting of von Neumann reduced entropy and concurrence (to study the atom-field entanglement) and negativity (to obtain the DEM between the atoms) has been numerically studied. The main results of the paper are concisely listed in what follows. \\
The numerical results of the von Neumann entropy as well as the concurrence showed that the existence of the Kerr medium intensively decreases the maximum amount of entanglement between atomic and field subsystems, while the detuning parameter has no remarkable effect on the maximum value of the DEM.
Looking deeply at the numerical results of von Neumann entropy and comparing them with the ones of concurrence implies the fact that, even though the general behavior of both of them are similar, for our system together with the considered conditions, the measure of concurrence is more sensitive than the von Neumann entropy.
The presented results of the negativity showed that the maximum value of the negativity is decreased due to the presence of detuning parameter. Also, the DEM between two atoms was destroyed, when the Kerr nonlinearity was considered.
 In an overall view, it was illustrated that the amount of DEM between subsystems can be tuned by suitably selecting the nonlinearity parameters related to the atom-field system.

 \vspace {2 cm}


 \end{document}